\documentclass[pss]{wiley2sp} % provides pss two-column style
\usepackage{amsmath}
\begin{document}

% Title of the article
\title{Ultra-low acoustic-phonon-limited
mobility and giant phonon-drag thermopower in MgZnO/ZnO
heterostructures}

% Abbreviated title for the page headers
\titlerunning{Ultra-low acoustic-phonon-limited mobility and giant phonon-drag thermopower}

% Authors
\author{%
  Margarita Tsaousidou\textsuperscript{\Ast,\textsf{\bfseries 1}}}

% Abbreviated list of authors for the page headers
\authorrunning{Margarita Tsaousidou}

%E-mail-address of corresponding author
\mail{e-mail
  \textsf{rtsaous@upatras.gr}}

% author's affiliations/addresses
\institute{%
  \textsuperscript{1}\,Materials Science Department, University of Patras, Patras 26 504, Greece}

\received{XXXX, revised XXXX, accepted XXXX} % do not change, will be filled in by the publisher
\published{XXXX} % do not change, will be filled in by the publisher

% Please select about four verbal keywords for your manuscript.
\keywords{MgZnO/ZnO heterostructures, electron-phonon coupling,
acoustic-phonon-limited mobility, phonon-drag thermopower}

\abstract{%
% This is a macro for the typesetting of two-column text in an
% abstract. It will typeset the two arguments in \abstcol{}{} as the
% left and right column inside the abstract box. At the
% columnbreak there will be always a columnbreak (\par), so both
% columns start with a new paragraph. No automatic column height
% balancing is done.
%
% If used with a \titlefigure it will silently output both
% parameters as consecutive paragraphs.
%
% The macro is defined exclusively inside the argument of \abstract{};
% if used outside it will raise an error.
%
%Usage: \abstcol{<left column>}{<right column>}
\abstcol{%
We present numerical simulations of the acoustic-phonon-limited
mobility, $\mu_{ac}$, and phonon-drag thermopower, $S^{g}$, in
two-dimensional electron gases confined in MgZnO/ZnO
heterostructures. The calculations are based on the Boltzmann
equation and are made for temperatures in the range 0.3-20~K and
sheet densities $0.5$-$30\times 10^{15}$~m$^{-2}$. The theoretical
estimations of $\mu_{ac}$ are in good agreement with the
experiment without any adjustable parameters. We find that the
magnitude of $\mu_{ac}$ }{%
is dramatically decreased in relation to GaAs based
heterostructures. The phonon-drag thermopower, $S^{g}$, which
according to Herring's expression is inversely proportional to
$\mu_{ac}$ is severely increased exceeding 200~mV/K at $T=5$~K
depending on sheet density. The giant values of $S^{g}$ lead to a
strong improvement of the figure of merit $ZT$ at low
temperatures. Our findings suggest that MgZnO/ZnO heterostructures
can be candidates for good thermoelectric materials at cryogenic
temperatures.}}

% The class file requires the standard graphicx Latex package. See the 'LaTeX
% standard graphics and color packages documentation' for more information at
% <http://tug.ctan.org/tex-archive/macros/latex/required/graphics/grfguide.pdf>.
%
% Accepted figure file formats depend on which LaTeX flavour is used.
% Classic LaTeX is always able to use Encapsulated Postscript (EPS);
% PDFLaTeX can't use this but accepts PDF, JPG, PNG, and GIF formats.
%
% See examples for implementing graphics in floating figure environments later in this file.
% If \titlefigure is given, it takes as its mandatory parameter the
% name (without extension) of some figure file.

\maketitle   % please do not remove

\section{Introduction}
In the last few years the transport properties of two-dimensional
electron gases (2DEGs) confined in MgZnO/ZnO heterostructures have
attracted a considerable amount of research interest. Recent
advances on growth techniques have enabled the realization of very
clean 2DEGs with mobilities reaching $8\times
10^{5}$~cm$^{2}$/Vs~\cite{Tsukazaki1,Falson,Maryenko} and the
observation of the integer~\cite{Tsukazaki2} and the fractional
quantum Hall effect~\cite{Tsukazaki1,Falson,Maryenko,Kozuka}. In
addition 2DEGs in MgZnO/ZnO heterostructures are ideal systems for
studying correlation effects~\cite{Kasahara} because the
interaction parameter $r_{s}=(\sqrt{\pi
n_{s}}\alpha^{*}_{B})^{-1}$ (with $n_{s}$ being the sheet density
and $\alpha^{*}_{B}$ the effective Bohr radius) is much larger
compared to the one in AlGaAs/GaAs heterostructures mainly due to
the larger effective mass in ZnO based materials.

In this Letter we study the effect of electron coupling with
acoustic phonons on the mobility and thermopower. So far
theoretical studies have been successful in interpreting mobility
data at low $T$ by introducing charged-impurity and
interface-roughness scattering~\cite{Gold1} but the higher
$T$-regime where acoustic phonons become important has remained
unexplored until now. Recently Falson {\em et al}~\cite{Falson}
presented a set of mobility data in MgZnO/ZnO heterostructures in
the temperature range 0.3-20~K and for sheet densities 0.7 to
$20\times 10^{15}$~m$^{-2}$. These data show clear evidence of
acoustic-phonon scattering at higher $T$. Here we present
numerical simulations of the acoustic-phonon-limited mobility,
$\mu_{ac}$, that are based on the semiclassical Boltzmann
equation~\cite{Tsaousidou1} and we obtain good agreement with the
mobility data of Ref.~\cite{Falson} without adjustable parameters.
It is found that $\mu_{ac}$ in MgZnO/ZnO heterostructures is
reduced by over two orders of magnitude in relation to GaAs based
quantum wells (QWs).

We also investigate, for the first time, the phonon-drag
thermopower, $S^{g}$, in 2DEGs confined in MgZnO/ZnO
heterostructures. $S^{g}$ is the contribution to thermopower that
arises due to the momentum exchange between electrons and
non-equilibrium acoustic phonons in the presence of a weak
in-plane $\nabla T$. Phonon-drag thermopower in 2DEGs has been
extensively studied in both a theoretical and an experimental
level~\cite{Gallagher,Tsaousidou}. According to Herring's
expression~\cite{Herring} $S^{g}$ is inversely proportional to
$\mu_{ac}$. Consequently in ZnO based 2D systems we expect a
dramatic increase of $S^{g}$ in relation to GaAs QWs. Namely, for
the samples of Ref.~\cite{Falson} we find that the magnitude of
$S^{g}$ exceeds 200 mV/K depending on sheet density. These are the
larger values that have been predicted so far for 2DEGs systems.
Due to the huge magnitude of $S^{g}$ a strong enhancement of $ZT$
is predicted at low $T$.

\section{Theory}
We assume that the 2DEG lies on the $xy$-plane. The
acoustic-phonon-limited mobility that is related to the scattering
of 2D electrons with wave vector ${\bf k}=(k_{x},k_{y})$ by 3D
acoustic phonons with wave vector ${\bf Q}=({\bf q},q_{z})$ is
obtained from~\cite{Tsaousidou1}
\begin{equation}
\mu_{ac}^{-1}=\frac{m^{*}}{e}\left\langle\frac{1}{\tau_{ac}}\right\rangle,
\end{equation}
where $m^{*}$ is the electron effective mass, $e$ is the magnitude
of the electron charge and $\langle1/\tau_{ac}\rangle$ is a
suitable average over the electron energy $\epsilon_{\bf k}$ of
the microscopic electron-phonon (e-ph ) scattering rate
$1/\tau_{ac}$ given by~\cite{Tsaousidou1}
\begin{equation}
\left<\frac{1}{\tau_{ac}}\right>=\frac{\int d\epsilon_{\bf
k}\epsilon_{\bf k}[df^{0}(\epsilon_{\bf k})/d\epsilon_{\bf
k}]\langle 1/\tau_{ac}(\epsilon_{\bf k})\rangle}{\int
d\epsilon_{\bf k}\epsilon_{\bf k}[df^{0}(\epsilon_{\bf
k})/d\epsilon_{\bf k}]},
\end{equation}
where $f^{0}(\epsilon_{\bf k})$ is the Fermi-Dirac function.
$1/\tau_{ac}$ is obtained by solving the Boltzmann equation in the
presence of a weak electric field in the relaxation time
approximation. Then for the average $\langle1/\tau_{ac}\rangle$ we
get~\cite{Tsaousidou1}
\begin{eqnarray}
\left<\frac{1}{\tau_{ac}}\right>=\frac{(2m^{*})^{1/2}}{2\pi
n_{s}\hbar^{2}k_{B}T}\sum_{\lambda,{\bf Q}}q\frac{|U_{{\bf
Q}\lambda}|^{2}}{\epsilon^{2}(q,T)}N^{0}_{{\bf Q}\lambda}Z(q_{z})\nonumber\\
\times \int_{\gamma}^{\infty}d\epsilon_{\bf
k}\frac{f^{0}(\epsilon_{\bf k})[1-f^{0}(\epsilon_{\bf
k}+\hbar\omega_{{\bf Q}\lambda})]}{\sqrt{\epsilon_{\bf
k}-\gamma}},
\end{eqnarray}
where $\lambda$ denotes the phonon mode (one longitudinal and two
transverse), $|U_{{\bf Q}\lambda}|^{2}$ is the square of the e-ph
coupling matrix elements, $\epsilon(q,T)$ is the static 2D
dielectric function, $\hbar\omega_{{\bf Q}\lambda}$ is the phonon
energy, $N_{{\bf Q}\lambda}^{0}$ is the phonon distribution in
equilibrium, $\gamma=(\hbar\omega_{{\bf
Q}\lambda}-\epsilon_{q})^{2}/4\epsilon_{q}$ (where
$\epsilon_{q}=\hbar^{2}q^{2}/2m^{*}$). Finally, $Z(q_{z})$ is the
form factor \begin{equation} Z(q_{z})=\left|\int
\phi^{*}_{0}(z)\exp(iq_{z}z)\phi_{0}(z)dz\right|^{2}
\end{equation}
where for the envelope function in the ground state
$\phi_{0}(z)$ we use the Fang-Howard~\cite{Fang} wave functions.

The square of the e-ph matrix elements $|U_{{\bf Q}\lambda}|^{2}$
is
\begin{equation}
|U_{{\bf Q}\lambda}|^{2}=\frac{\hbar Q^{2}\Xi^{2}_{Eff}({\bf
Q}\lambda)}{2\rho V\omega_{{\bf Q}\lambda}}
\end{equation}
where $\rho$ and $V$ are, respectively, the density and the volume
of the sample. The term $\Xi_{Eff}({\bf Q}\lambda)$ denotes the
'effective' acoustic potential describing the e-ph coupling and
accounts for both the deformation potential and the piezoelectric
coupling. In ZnO the conduction band is isotropic and only
longitudinal acoustic phonons are coupled with electrons via
deformation potential coupling. In this case the deformation
potential is described by a single constant $\Xi_{d}$. In
semiconductors with wurtzite structure, such as ZnO, the
contribution to $\Xi^{2}_{Eff}({\bf Q}\lambda)$ due to
piezoelectric e-ph coupling is~\cite{Krummheuer}
\begin{eqnarray}
\Xi^{2}_{Piez}({\bf
Q}\lambda)&=&\frac{e^{2}}{Q^{6}}\{h_{15}(q_{x}^{2}+q_{y}^{2})({\bf
e}_{{\bf Q}\lambda})_{z}+h_{33}q_{z}^{2}({\bf e}_{{\bf
Q},\lambda})_{z}\nonumber\\
&+&(h_{15}+h_{31})q_{z}[q_{x}({\bf e}_{{\bf
Q}\lambda})_{x}+q_{y}({\bf e}_{{\bf Q}\lambda})_{y}]\}^{2},\nonumber\\
\end{eqnarray}
where $({\bf e}_{{\bf Q}\lambda})_{i}$ is the $i$-component of the
phonon polarization vector and $h_{15}$, $h_{33}$, $h_{31}$ are
the non-zero elements of the piezoelectric tensor.

The 2D dielectric function $\epsilon(q,T)$ has the
form~\cite{Calmels,Ando}
\begin{equation}
\epsilon(q,T)=1+\frac{e^{2}}{2\epsilon_{0}\epsilon_{r}q}\Pi(q,T)F(q)\xi(q)[1-G(q)],
\end{equation}
where $\epsilon_{0}$ is the permittivity of vacuum, $\epsilon_{r}$
is the relative permittivity of ZnO, $\Pi(q,T)$ is the
polarisability~\cite{Maldague}, $F(q)$ is the screening form
factor that accounts for the finite extent of the electron wave
function in the $z$-direction~\cite{Ando}, $\xi(q)=1$ when
$q/2k_{F}<1$ and $\xi(q)=1-\sqrt{1-(2k_{F}/q)^{2}}$ when
$q/2k_{F}\geq 1$ (with $k_{F}$ being the Fermi wave number).
$G(q)$ is the local-field correction (LFC) factor that describes
exchange and correlation effects beyond the random-phase
approximation. It is given by~\cite{Calmels}
\begin{equation}
G(x)=r_{s}^{2/3}\frac{1.402x}{\sqrt{2.644C_{12}^{2}+x^{2}C_{22}^{2}}},
\end{equation}
where $x=q/q_{0}$ and $q_{0}=2/(r_{s}^{2/3}\alpha^{*}_{B})$.
$C_{12}$ and $C_{22}$ depend on $r_{s}$ and their expressions are
given in Ref.~\cite{Calmels}. We note that the LFC becomes
important at low densities where the interaction parameter $r_{s}$
becomes large.

The phonon-drag thermopower is related to $\mu_{ac}$ via Herring's
expression~\cite{Herring}. Although Herring's expression was
phenomenological when it was first proposed, later
studies~\cite{Tsaousidou1,Miele} based on the semiclassical
Boltzmann framework showed that this expression was precise in
2DEGs. We can write~\cite{Tsaousidou1,Miele}
\begin{equation}
S^{g}=-\sum_{\lambda}\frac{v_{s}^{\lambda}l_{ph}m^{*}}{eT}\left\langle\frac{1}{\tau_{ac}^{\lambda}}\right\rangle
=-\sum_{\lambda}\frac{v_{s}^{\lambda}l_{ph}}{T}\left\langle\frac{1}{\mu_{ac}^{\lambda}}\right\rangle,
\end{equation}
where $v_{s}^{\lambda}$ is the sound velocity for the $\lambda$
mode and $l_{ph}$ is the phonon-mean-free path which is assumed to
be $\lambda$-independent. At low temperatures $l_{ph}$ is
determined by boundary phonon scattering and depends only on the
dimensions of the sample. Detailed numerical calculations showed
that the average $\langle 1/\mu_{ac}\rangle$ describes accurately
the inverse of acoustic-phonon-limited mobility.

\section{Numerical simulations and comparison with the
experiment} The values for the parameters used in the calculations
are $m^{*}=0.29~m_{e}$ (see, for example, Ref.~\cite{Gold1}),
$v_{s}^{L}=6167$~m/s and $v_{s}^{T}=2561$~m/s~\cite{Sarasamak},
$\epsilon_{r}=8.5$~\cite{Gold1}, $\rho=6.1\times
10^{3}$~Kg/m$^{3}$~\cite{Look}, and $\Xi_{d}=15$~eV~\cite{Look},
$h_{15}=-6.4\times 10^{9}$~V/m, $h_{31}=-7.6\times 10^{9}$~V/m,
and $h_{33}=17.5\times 10^{9}$~V/m~\cite{Auld}.

In Fig.1 the dots are the experimental data~\cite{Falson} for the
total mobility in four samples with sheet densities $1.4$, $1.7$,
$4.5$ and $7\times 10^{15}$~m$^{-2}$. The red dashed lines are the
numerical simulations of $\mu_{ac}$ based on Eqs.~(1)-(8). The
total mobility is obtained by
$\mu_{tot}^{-1}=\mu_{ac}^{-1}+\mu_{el}^{-1}$ where $\mu_{el}$ is
the mobility limited by elastic scattering ($\mu_{el}$ is obtained
from the data at $T=0.3$~K). The calculations of $\mu_{tot}$ are
shown as blue solid lines. The agreement with the experiment is
good without adjustable parameters. Good agreement was found also
for another sample with $n_{s}=20\times 10^{15}$~m$^{-2}$ while
for a very dilute sample with $n_{s}=0.68\times 10^{15}$~m$^{-2}$
the theory underestimates the data by a factor of two at the
highest $T$ examined (not shown here).

\begin{figure}[b]%
\includegraphics*[angle=-90, width=8.2cm]{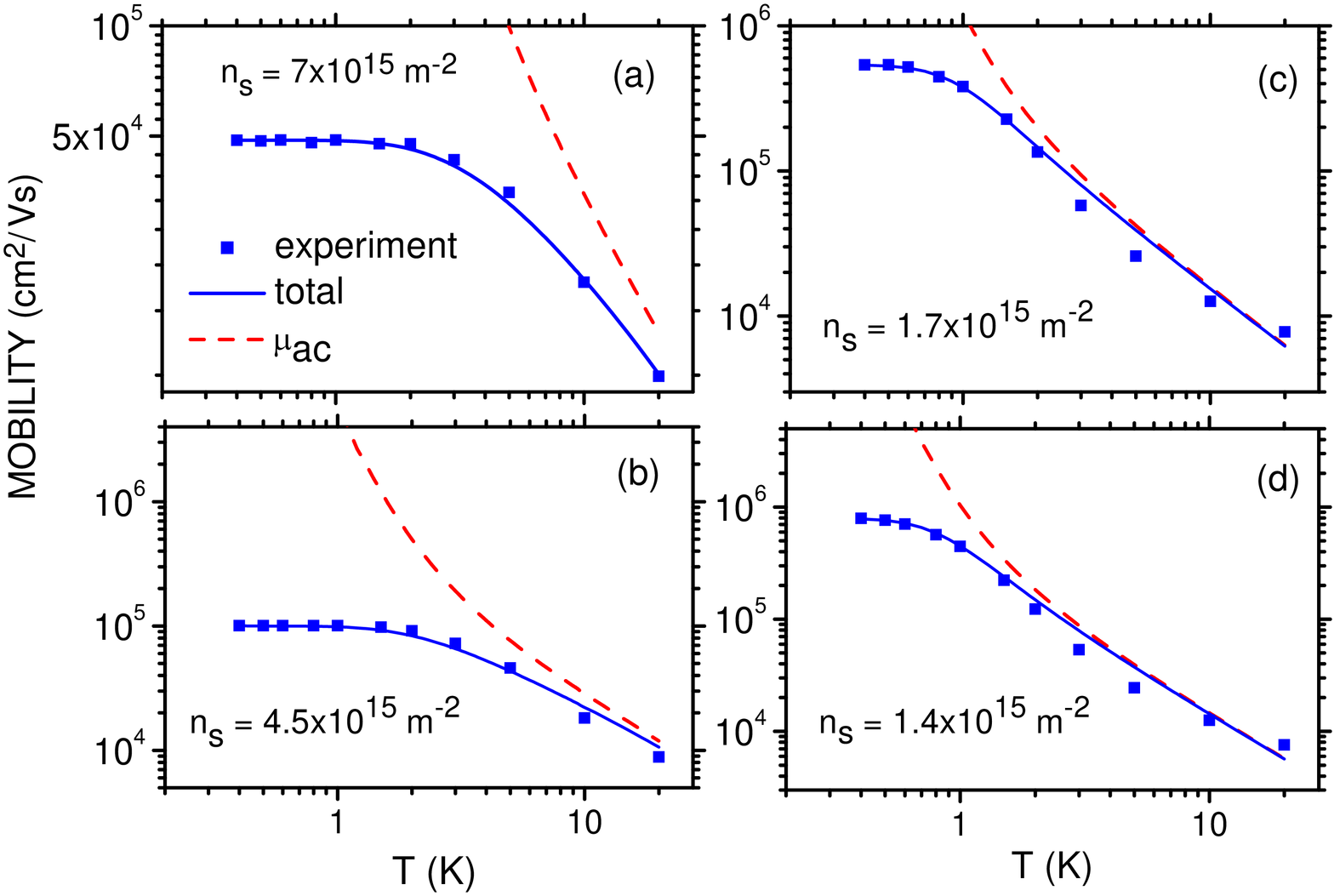}
\caption{Mobility as a function of temperature for four samples
with sheet densities 1.4, 1.7, 4.5, and $7\times
10^{15}$~m$^{-2}$. The blue solid and the red dashed lines denote
the calculations of the $\mu_{tot}$ and $\mu_{ac}$, respectively.
The symbols are the experimental data [2].}
\end{figure}

The effect of the LFC term $G(q)$ in the dielectric function is
shown in Fig.2 for a dilute sample with $n_{s}=1.4\times
10^{15}$~m$^{-2}$ ($r_{s}=9.7$). The calculations of $\mu_{ac}$
without the incorporation of the $G(q)$ term in Eq.~(7) are
enhanced by over a factor of three (green dashed-dotted line). In
Fig.2 we show also the calculations of $\mu_{ac}$ for a 2DEG
confined in an AlGaAs/GaAs heterostructure (black dotted line)
with the same sheet density. We see that $\mu_{ac}$ in MgZnO/ZnO
heterostructures is severely suppressed in relation to GaAs/AlGaAs
counterparts.

\begin{figure}[t]%
\includegraphics*[angle=-90, width=8.2cm]{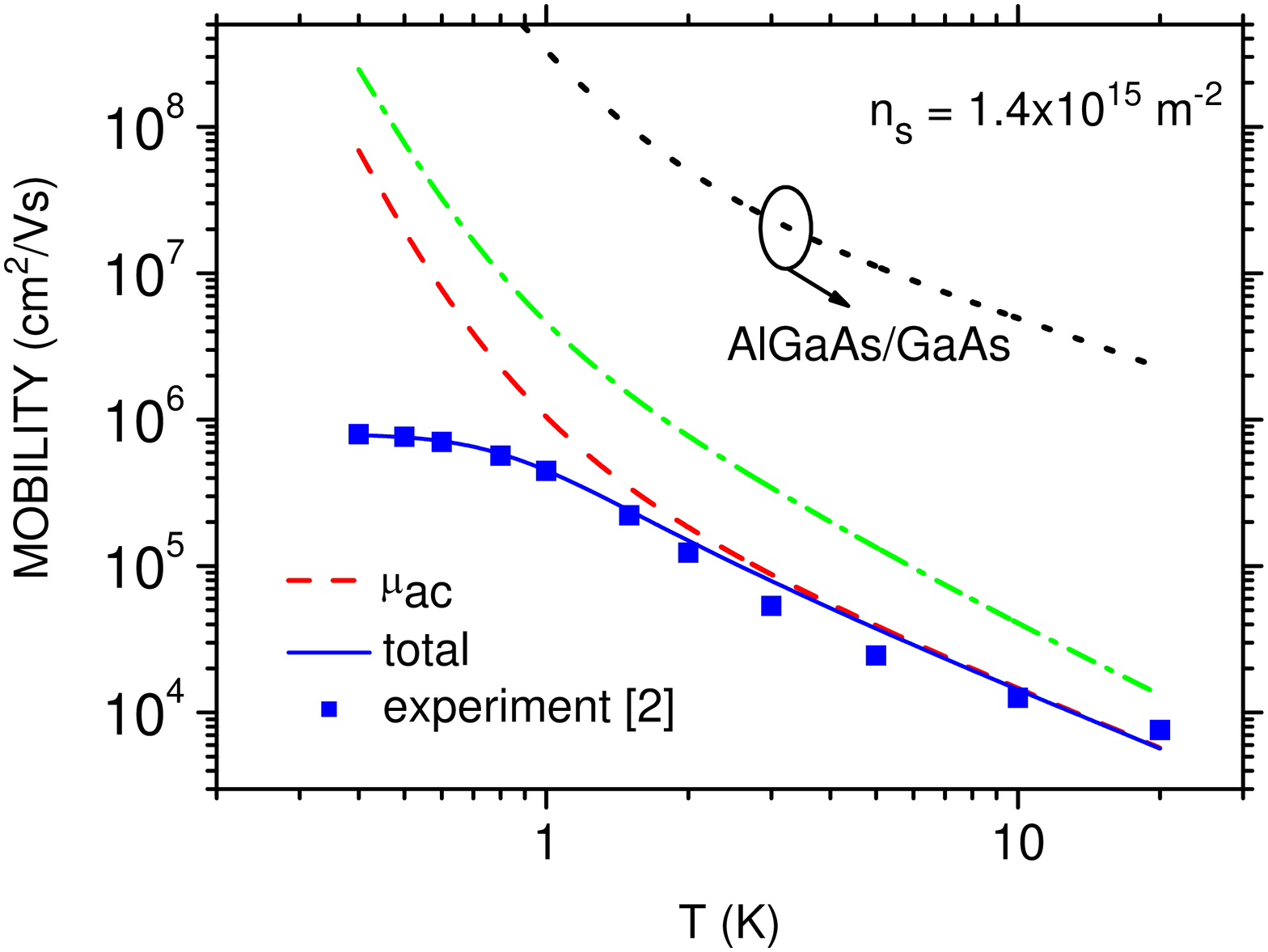}
\caption{Mobility as a function of temperature for the sample with
$n_{s}=1.4\times 10^{15}$~m$^{-2}$. The green dashed-dotted line
shows the results for $\mu_{ac}$ when the local-field correction
is ignored. The calculation of $\mu_{ac}$ for an AlGaAs/GaAs
heterostructure with the same $n_{s}$ is shown as black dotted
line.}
\end{figure}

\begin{figure}[b]%
\includegraphics*[width=0.8\linewidth,height=\linewidth]{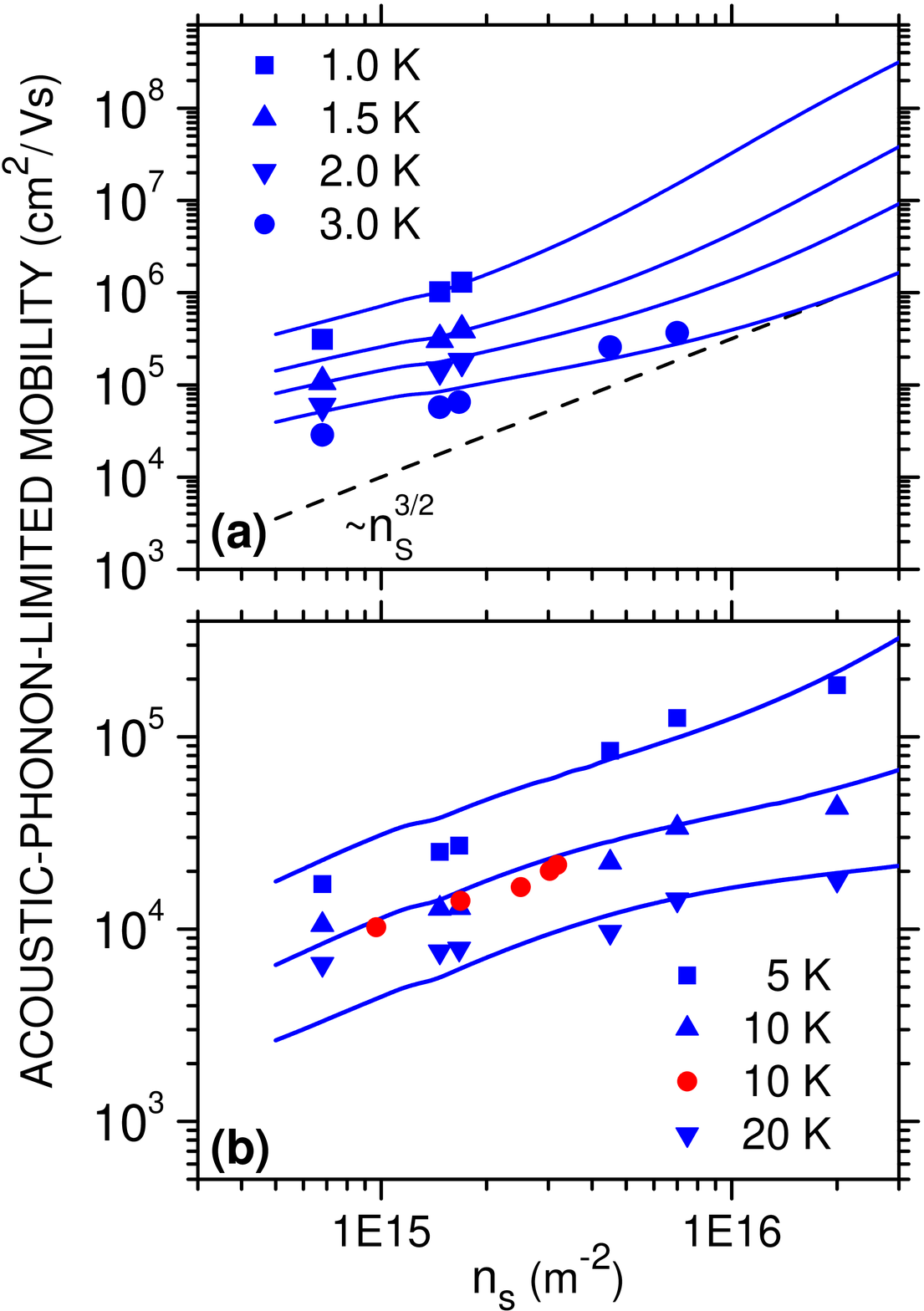}
\caption{%
Acoustic-phonon-limited mobility as a function of sheet density
for $T$ in the range 1-20~K. The solid lines are the numerical
simulations and the symbols are the experimental data from Ref.[2]
(blue) and Ref.[1] (red).}
\end{figure}

The dependence of $\mu_{ac}$ on sheet density is shown in Fig.3.
Inspection of Eq.(3) shows that at low $T$ ($q\ll 2k_{F}$) and for
degenerate 2DEGs $\mu_{ac}$ varies as $n_{s}^{3/2}$. The deviation
observed at low $n_{s}$ is due to the breakdown of the low-$T$
approximations. The elastic scattering is strong at low
temperatures for the samples with densities higher than $1.7\times
10^{15}$~m$^{-2}$ and we cannot identify $\mu_{ac}$ accurately.
For this reason we do not present the relevant data in Fig.3a.

In Fig.4a we show the calculations of -$S^{g}$ by using
Eqs.(3)-(9) for three 2DEGs confined in MgZnO/ZnO heterostructures
with sheet densities 0.7 (blue line), 1.4 (red line) and $7\times
10^{15}$~m$^{-2}$ (green line). $l_{ph}$ is taken to be 1~mm. At
low $T$ $S^{g}$ follows a $T^{4}$ law which is characteristic for
piezoelectric e-ph coupling~\cite{Tsaousidou}. In the same figure
we present also the calculations of -$S^{g}$ for a 2DEG in an
AlGaAs/GaAs heterostructure with $n_{s}=0.7\times
10^{15}$~m$^{-2}$. Quite remarkably the magnitude of phonon drag
is dramatically enhanced in MgZnO/ZnO heterostructures exceeding
200~mV/K at $T=5$~K for low $n_{s}$. This huge increase is due to
the following characteristics of ZnO based 2D structures: (i) the
large effective mass, (ii) the strong piezoelectric e-ph coupling,
and (iii) the decrease of the screening effects due to exchange
and correlation effects. (The same characteristics are responsible
for the ultra-low $\mu_{ac}$.)

\begin{figure}[t]%
\includegraphics*[width=0.8\linewidth,height=\linewidth]{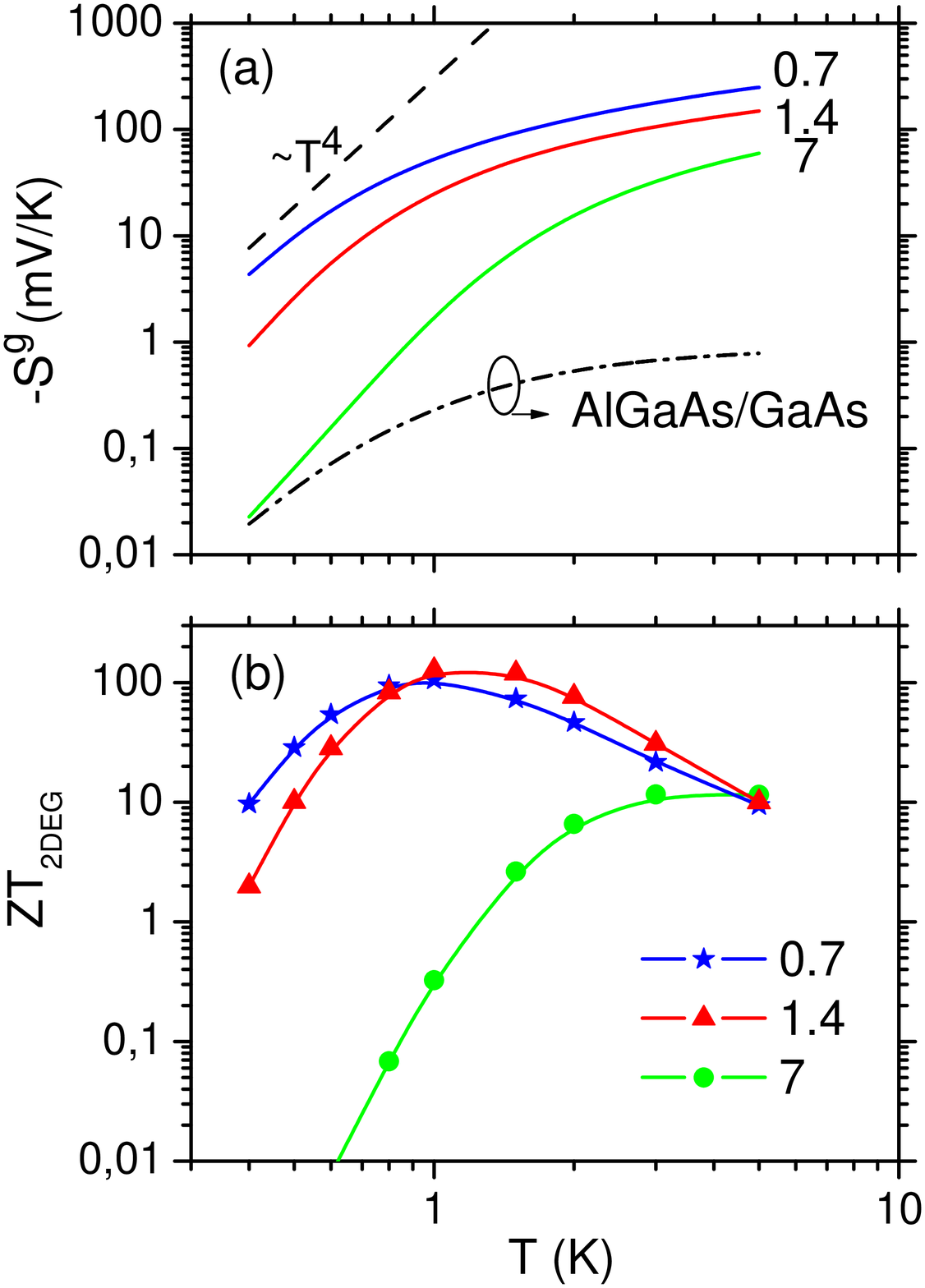}
\caption{-$S^{g}$ (a) and $ZT_{2DEG}$ (b) as a function of $T$ for
three 2DEGs in a MgZnO/ZnO heterointerface with $n_{s}= 0.7, 1.4$
and $7\times 10^{15}$~m$^{-2}$. In (a) the dashed-dotted line
refers to a 2DEG in an AlGaAs/GaAs heterostructure with
$n_{s}=0.7\times 10^{15}$~m$^{-2}$.}
\end{figure}

We note that at low $T$ the diffusion thermopower overwhelms
$S^{g}$. For the 2DEGs in a MgZnO/ZnO heterostructure shown in
Fig.~4 the crossover occurs at $T$ in the range 0.1-0.2~K assuming
small-angle electron scattering.

The giant -$S^{g}$ results to a significant enhancement of the
thermoelectric efficiency at low $T$. The calculations of the
figure of merit $ZT_{2DEG}$ for the 2DEGs confined in MgZnO/ZnO
heterostructures with $n_{s}=0.7, 1.4$, and $7\times
10^{15}$~m$^{-2}$ are shown in Fig.4b. More analytically for
$ZT_{2DEG}$ we use the expression $ZT_{2DEG}=(S^{g})^{2}\sigma
T/\kappa_{ph}$ where $\sigma=n_{s}e\mu_{tot}/w$ with $w$ being the
thickness of the 2DEG (we assume that $w=10$~nm) and
$\kappa_{ph}=4860\,l_{ph} T^{3}$ is the phonon conductivity in the
Debye approximation. We note that $ZT_{2DEG}$ is proportional to
$l_{ph}$. At low $T$ $l_{ph}$ is specified by the dimensions of
the sample and can be a few mm. Thus $ZT_{2DEG}$ can be even
larger that what shown in Fig.4b where $l_{ph}$ is taken to be
1~mm. In the calculation of the figure of merit we use the
thickness of a single layer. The conductivity of the barrier layer
is smaller than that of the 2DEG layer. The effective conductivity
is $\sigma_{eff}=\sigma/(1+N_{barrier})$~\cite{Ohta} where
$N_{barrier}$ is the number of the unit cells of the barrier
layer. Consequently, $ZT_{2DEG}$ is decreased by the factor
$1+N_{barrier}$.

\section{Conclusions}
In summary, we predict an ultra-low acoustic-phonon-limited
mobility and a giant phonon-drag thermopower in MgZnO/ZnO
heterostructures. The theoretical estimates of $\mu_{ac}$ are in
very good agreement with the experiment without adjustable
parameters. This gives us confidence about our understanding of
the mechanisms of e-ph coupling in the present system and the
accuracy of our calculations for $S^{g}$. We find that at low $T$
the magnitude of $S^{g}$ can exceed 200~mV/K. This value of
$S^{g}$ is the largest ever predicted in 2DEGs. Finally, we
predict a dramatic increase of the effective figure of merit that
depends on the sheet density, the phonon-mean-free path and the
number of barrier layers.

\begin{acknowledgement}
The author wishes to thank Dr. Falson for providing the mobility
data of Ref.[2] appearing in Fig.1, Fig.2, and Fig.3.
\end{acknowledgement}


\begin{thebibliography}{[1]}

\bibitem{Tsukazaki1}%
 A.~Tsukazaki {\em et al.}, Nature Mater. \textbf{9}, 889 (2010).

\bibitem{Falson}%
 J.~Falson {\em et al.}, Appl. Phys. Express \textbf{4}, 091101 (2011).

\bibitem{Maryenko}%
D.~Maryenko {\em et al.}, Phys. Rev. Lett. \textbf{108}, 186803
(2012).

\bibitem{Tsukazaki2}%
 A.~Tsukazaki {\em et al.}, Science \textbf{315}, 1388 (2007).

\bibitem{Kozuka}%
 Y.~Kozuka {\em et al.}, Phys. Rev. B \textbf{85}, 075302 (2012).

\bibitem{Kasahara}%
Y.~Kasahara {\em et al.}, Phys. Rev. Lett. \textbf{109}, 246401
(2012).

\bibitem{Gold1}%
A.Gold, Appl. Phys. Lett. \textbf{96}, 242111 (2010); A. Gold, J.
Appl. Phys. \textbf{110}, 043702 (2011).

\bibitem{Tsaousidou1}%
M.~Tsaousidou {\em et al.}, Phys. Rev. B \textbf{64}, 165304
(2001).

\bibitem{Gallagher}%
B.\,L. Gallagher and P.\,N. Butcher, in Handbook on
Semiconductors, edited by P.\,T. Landsberg (Elsevier, Amsterdam,
1992), vol. 1, p.\,817; R.~Fletcher, E. Zaremba, and U. Zeitler,
in Electron-Phonon Interactions in Low-Dimensional Structures,
edited by L.~Challis (Oxford Science Publications, Oxford, 2003),
p.\,149.

\bibitem{Tsaousidou}%
M.~Tsaousidou, in The Oxford Handbook of Nanoscience and
Technology, edited by A.\,V. Narlikar and Y.\,Y. Fu (Oxford
University Press, Oxford, 2010), vol. II, p.\,477.

\bibitem{Herring}%
C.~Herring, Phys. Rev. \textbf{96}, 1163 (1954).

\bibitem{Fang}%
F.\,F. Fang and W.\,E. Howard, Phys. Rev. Lett. \textbf{16}, 797
(1966).

\bibitem{Krummheuer} B.~Krummheuer {\em et al.}, Phys. Rev. B {\bf
71}, 235329 (2005).

\bibitem{Calmels}%
A.~Gold and L. Calmels, Phys. Rev. B \textbf{48}, 11622 (1993).

\bibitem{Ando}%
T.~Ando {\em et al.}, Rev. Mod. Phys. \textbf{54}, 437 (1982).

\bibitem{Maldague}%
P.\,F. Maldague, Surf. Sci. \textbf{73}, 296 (1978).

\bibitem{Miele}%
A.~Miele {\em et al.}, Phys. Rev. B \textbf{58}, 13181 (1998).

\bibitem{Sarasamak}%
K.~Sarasamak {\em et al.}, Phys. Rev. B \textbf{82}, 035201
(2010).

\bibitem{Look}%
D.\,C. Look, Sem. Sci. Tech. {\bf 20}, S55 (2005).

\bibitem{Auld}%
B.\,A. Auld, in Acoustic Fields and Waves in Solids, (John Wiley
and Sons, USA, 1973), vol. I, p.\,378.

\bibitem{Ohta}%
H.~Ohta {\em et al.}, Nature Mater. {\bf 6}, 129 (2007).




\end{thebibliography}
\end{document}